\input{aipcheck}

\documentclass[
    ,final            
  ]
  {aipproc}

\layoutstyle{6x9}

\begin{document}

\title{Fusion Hindrance and the Role of Shell Effects in the Superheavy
Mass Region}


\author{Y.~Aritomo}{
  address={Flerov Laboratory of Nuclear
Reactions, JINR, Dubna, Russia} }


\begin{abstract}
 We present the first attempt of the systematical investigation about
the effects of shell correction energy for dynamical processes,
which include fusion, fusion-fission and quasi-fission processes. In
the superheavy mass region, for the fusion process, the shell
correction energy plays a very important role and enhances the
fusion probability, when the colliding partner has strong shell
structure. By analyzing the trajectory in the three-dimensional
coordinate space with a Langevin equation, we reveal the mechanism
of the enhancement of the fusion probability caused by shell
effects.
\end{abstract}

\maketitle


\section{Introduction}

\hspace*{3.5mm} In the heavy ion fusion reaction, with increasing
atomic numbers of the target and the projectile, it becomes more
difficult to make a compound nucleus due to the strong Coulomb
repulsion force and strong dissipation force. In the superheavy mass
region, this difficulty is more remarkable. Though the mechanism of
fusion-fission reaction in the heavy mass-region is not clear,
generally we recognize the existence of fusion hindrance.

The fusion hindrance is mainly caused by the macroscopic properties
of the colliding partner. On the contrary, the fusion is enhanced by
the shell structure of the nuclei \cite{moll97,armb99}. In the
experiments, especially cold fusion reaction, these advantages are
used to synthesize the superheavy nuclei \cite{hofm00,mori03}. To
understand the fusion mechanism clearly, it is better to treat
separately the fusion hindrance and the fusion enhancement, that is
to say, the macroscopic aspect and the microscopic one.

In our previous study \cite{ari05I}, we have discussed the fusion
hindrance in the superheavy mass region. We presented the origin of
the fusion hindrance systematically by the trajectory calculation.
In the present paper, as fusion enhancement, we focus on the
influence of shell effects. It is known that the nuclear structure
of the projectile-target combinations correlates to the touching
probability \cite{sato02,ikez03}, but it influences also the
dynamics from the touching point to the compound nucleus. We
investigate precisely how the trajectory behavior is influenced by
the shell correction energy.

\begin{figure}
\centerline{
\includegraphics[height=.33\textheight]{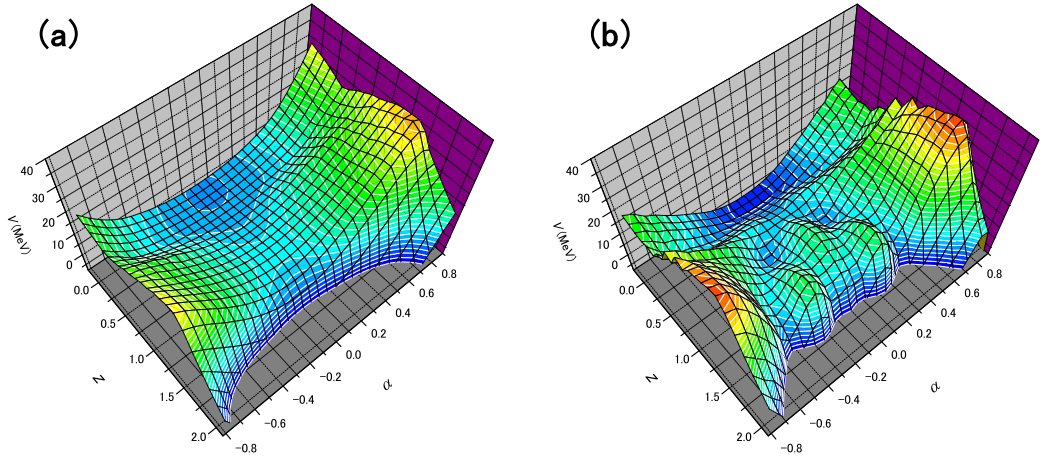}}
  \caption{Potential energy of the liquid drop model $V_{LD}$ (a) and with shell correction
   energy $V_{LD}+E^{0}_{shell}$ (b) for
$^{292}114$ in the $z-\alpha$ space $(\delta=0)$. The calculation is
done by a two-center shell model code \cite{suek74,iwam76}.}
\label{1-3dim}
\end{figure}

\section{Model}

\hspace*{3.5mm}  Using the same procedure as described in reference
\cite{ari04}, we investigate the dynamical processes. The
fluctuation-dissipation model with the Langevin equation is
employed. We adopt the three-dimensional nuclear deformation space
given by two-center parameterization \cite{maru72,sato78}. The three
collective parameters involved in the Langevin equation are as
follows: $z_{0}$ (distance between two potential centers), $\delta$
(deformation of fragments) and $\alpha$ (mass asymmetry of the
colliding nuclei); $\alpha=(A_{1}-A_{2})/(A_{1}+A_{2})$, where
$A_{1}$ and $A_{2}$ denote the mass numbers of the target and the
projectile, respectively.

The multidimensional Langevin equation is given as
\begin{eqnarray}
\frac{dq_{i}}{dt}&=&\left(m^{-1}\right)_{ij}p_{j},\nonumber\\
\frac{dp_{i}}{dt}&=&-\frac{\partial V}{dq_{i}}
                 -\frac{1}{2}\frac{\partial}{\partial q_{i}}
                   \left(m^{-1}\right)_{jk}p_{j}p_{k}
                  -\gamma_{ij}\left(m^{-1}\right)_{jk}p_{k}\nonumber
                  +g_{ij}R_{j}(t),\\
\end{eqnarray}
where a summation over repeated indices is assumed. $q_{i}$ denotes
the deformation coordinate. $p_{i}$ is the conjugate momentum of
$q_{i}$. $V$ is the potential energy, and $m_{ij}$ and $\gamma_{ij}$
are the shape-dependent collective inertia parameter and dissipation
tensor, respectively. A hydrodynamical inertia tensor is adopted in
the Werner-Wheeler approximation for the velocity field, and the
wall-and-window one-body dissipation is adopted for the dissipation
tensor. Details are explained in reference \cite{ari04}.


\begin{figure}
\centerline{
\includegraphics[height=.58\textheight]{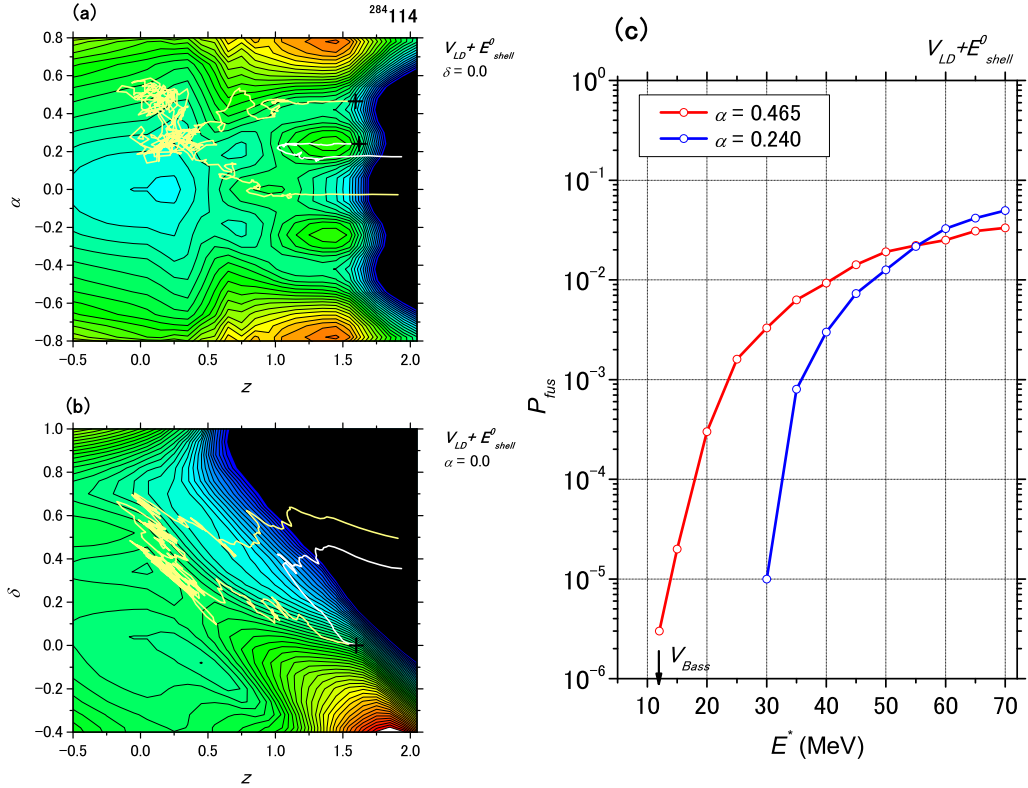}}
  \caption{Sample of the trajectory projected onto the $z-\alpha$ plane
at $\delta=0.0$ (a) and the $z-\delta$ plane at $\alpha=0.0$ (b) of
$V_{LD}+E^{0}_{shell}$ for $^{284}$114. The light yellow and white
lines denote the trajectories which start at $\alpha=0.46$ and 0.24
at $E^{*}=20$ MeV, respectively. Symbols are given in the text. (c)
Fusion probability with initial value $\alpha=0.46$ and
$\alpha=0.24$, which are denoted by the red and blue lines,
respectively.}
\end{figure}

\section{Results}

\subsection{Effect of the Cold Fusion Valleys in Fusion Process}

Figure \ref{1-3dim} shows the potential energy surface of the liquid
drop model (a) and with shell correction energy (b) for $^{292}114$
in the $z-\alpha$ space $(\delta=0)$, which is calculated by the
two-center shell model code \cite{suek74,iwam76}. When we consider
the shell correction energy, we can see the pronounced valleys which
lead to the compound nucleus. The valleys are called 'cold fusion
valleys' \cite{gra99,gra00}. It is said that these valleys enhance
the fusion probability. We discuss the effect of the cold fusion
valleys in the dynamical process using the trajectory calculation.


\begin{figure}
\includegraphics[height=0.33\textheight]{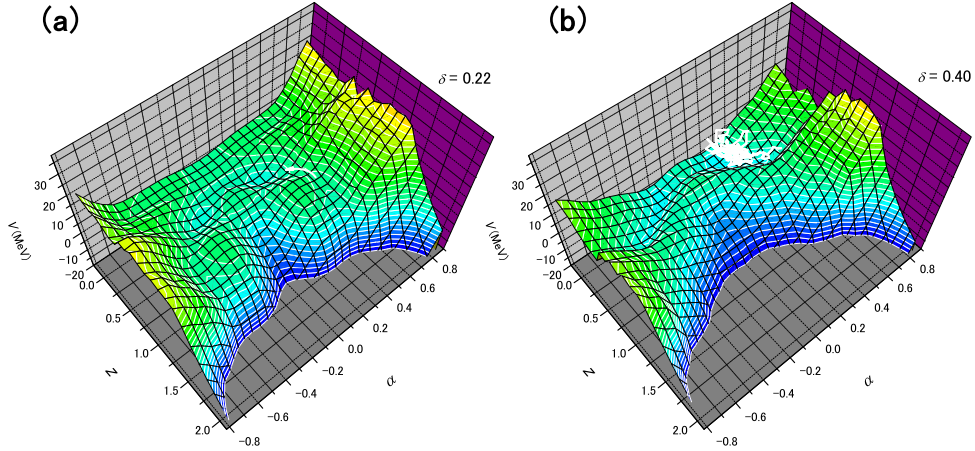}
  \caption{Potential energy surface $V_{LD}+E^{0}_{shell}$ in the $z-\alpha$ space
at (a) $\delta=0.22$ and (b) 0.40 for $^{284}$114, with a sample
trajectory.}
\end{figure}


As our first attempt, in the dynamical calculation, we employ the
potential energy of the liquid drop model with the full shell
correction energy, which corresponds to the potential energy surface
at the nuclear temperature $T=0$ MeV. It is represented by
$V_{LD}+V_{SH}(q,T=0)$, or $V_{LD}+E^{0}_{shell}$. We compare the
fusion probability with the different injection points, where the
shell correction energy has remarkably negative and positive values
like $\alpha=0.46$ and $0.24$, that correspond to the reaction
$^{76}$Ge + $^{208}$Pb and $^{108}$Ru + $^{176}$Yb, respectively.

Figure~2(a) shows the sample trajectories which are projected onto
the $z-\alpha$ $(\delta=0)$ plane of $^{284}114$. The contact point
is marked by $(+)$. The light yellow line denotes the trajectory
with the starting point $\alpha=0.46$ at the incident energy which
corresponds to the excitation energy of the compound nucleus
$E^{*}$=20 MeV. This trajectory looks to move along the valley till
$z \sim 1.0$. Then it enters the region near $z \sim 0.0$. On the
other hand, the trajectory with the starting point $\alpha=0.24$
goes till $z \sim 1.0$ without changing the mass asymmetry parameter
$\alpha$, which is denoted by the white line. It seems that the
trajectory overcomes the mountain located at $z=1.4, \alpha=0.24$.
We project these trajectories onto the $z-\delta$ plane at
$\alpha=0$ in Fig.~2(b). The trajectory with the starting point
$\alpha=0.46$ can enter around the compact shape region, but it is
not trapped by the pocket around the ground state. The trajectory
with the starting point $\alpha=0.24$ moves quickly in the $+\delta$
direction and goes to the fission region. We can say, the former is
a deep quasi-fission process (DQF) and the latter is a quasi-fission
process (QF) \cite{ari04,ari05I}.

Figure~2(c) shows the fusion probability with the initial value
$\alpha=0.46$ and $\alpha=0.24$, which are denoted by the red and
blue lines, respectively. The potential energy at the contact point
with $\alpha=0.46$ is lower than that with $\alpha= 0.24$, due to
the shell correction energy. Therefore, at low excitation energy,
the former fusion probability is larger than the latter one, because
in the former case the available kinetic energy at the contact point
is larger than in the latter case. The arrow denotes the Coulomb
barrier \cite{bass74}.


\begin{figure}
\includegraphics[height=0.60\textheight]{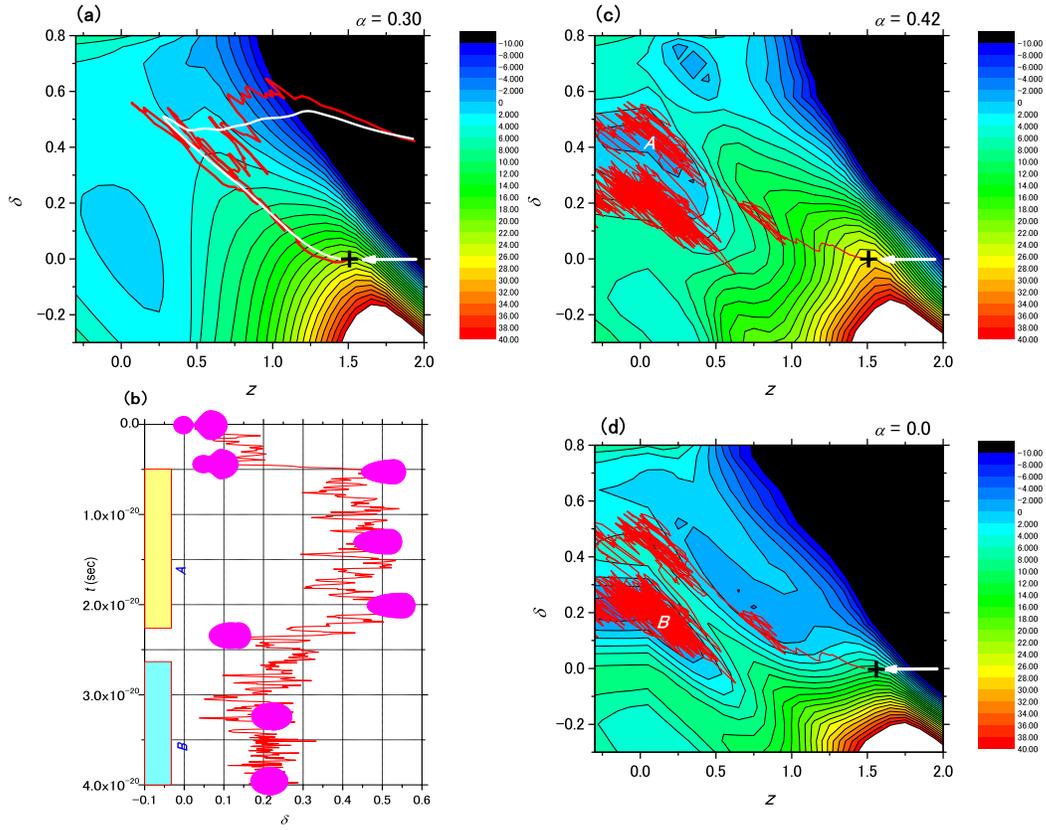}
  \caption{Sample of the trajectory projected onto the $z-\delta$ plane
at $\alpha$ which corresponds to the turning point. (a) $V_{LD}$
case and (b) $V_{LD}+E^{0}_{shell}$ case for $^{256}$No.
$V_{LD}+E^{0}_{shell}$ with $\alpha=0.0$ is denoted in (d). The
evolution of nuclear shapes is presented in (b).}
\end{figure}

\subsection{Evolution of the Cold Fusion Valleys in the Dynamical Process}

Usually the cold fusion valleys are discussed using the potential
energy surface on the $z-\alpha$ plane with $\delta=0$, like as
Fig.~1(b). However, in the dynamical process, the trajectory moves
in the large $+\delta$ direction. We should discuss the cold fusion
valleys with the dynamical evolution of the $\delta$ parameter.  For
the $\delta= 0$ case, which corresponds to Fig.~1(b), we can see the
remarkable the cold fusion valleys. With changing $\delta$ value in
the dynamical process, the cold fusion valleys also changes, which
is shown in Fig.~3. When the $\delta$ changes about 0.4, the cold
fusion valleys disappear.

\subsection{The Role of Shell Effects in the Fusion Process}

With the shell correction energy, we discuss another important
effect to enhance the fusion probability. In the discussion on
fusion hindrance \cite{ari05I}, we indicated that the turning point
is important. Fig.~4(a) shows the potential energy surface of
$V_{LD}$ for $Z$=102 at the turning point on the $z-\delta$ plane,
that is to say, $\alpha$ corresponds to the value at the turning
point. The mean trajectory is denoted by the white line. The gray
line denotes the trajectory with taking into account the
fluctuation. At the turning point $(z \sim 0.3, \delta \sim 0.5)$,
the trajectory moves to the fission direction due to the potential
landscape.

On the other hand, when we take into account the shell correction
energy in Fig.~4(c), the temporary pocket appears at the turning
point (indicated by $A$). At $\alpha=0$ in Fig.~4(d), the pocket
(indicated by $B$) corresponds to the ground state. When we take
into account the shell correction energy, the trajectory is trapped
in the pocket $A$ at the turning point, and is blocked from going to
the fission area \cite{ari05}. During the stay in the pocket $A$,
the mass asymmetry is relaxed. With appearing of the large pocket
$B$ at $\alpha \sim 0$, the trajectory moves to the pocket $B$. The
pocket $A$ helps the trajectory to enter the region corresponding to
the compound nucleus. Fig.~4(b) shows the time evolution of the
nuclear shape. The horizontal axis denotes the $\delta$. The
trajectory moves in the large $\delta$ direction quickly, then it is
trapped in the pocket $A$. During the stay in the pocket $A$, the
mass asymmetry changes. With approaching $\alpha=0$ the trajectory
moves into the pocket $B$. It corresponds to the compound nucleus.

The temporary pocket that appeared due to the shell correction
energy enhances the fusion probability. As the further study, we
take into account the temperature dependence of shell correction
energy.

\vspace{0.3cm}

The author is grateful to Professor M.~Ohta for his helpful
suggestions and valuable discussion throughout the present work.





\bibliographystyle{aipproc}   

\bibliography{sample}

\IfFileExists{\jobname.bbl}{}
 {\typeout{}
  \typeout{******************************************}
  \typeout{** Please run "bibtex \jobname" to optain}
  \typeout{** the bibliography and then re-run LaTeX}
  \typeout{** twice to fix the references!}
  \typeout{******************************************}
  \typeout{}
 }

\end{document}